\documentclass[aps,prb,groupedaddress,twocolumn]{revtex4}
\usepackage{epsfig}
\usepackage[dvipsnames,usenames]{color}
\usepackage{color}
\usepackage{amsmath}
\usepackage{comment}
\usepackage{array}
\usepackage{multirow}
\usepackage{tabularx}
\usepackage{float}
\usepackage{hyperref}

\emergencystretch=\maxdimen
\begin{document}
\title{Enhanced antiferromagnetic ordering tendency in staggered periodic Anderson model}
\author{Mi Jiang}
\affiliation{Stewart Blusson Quantum Matter Institute, University of British Columbia, Vancouver, BC, Canada}

\begin{abstract}
Heavy fermion compounds consisting of two or more inequivalent local moment sites per unit cell have been a promising platform of investigating the interplay between distinct Kondo screenings that is absent in the conventional systems containing only one rare-earth ion per unit cell.
We report a remarkable enhancement of the antiferromagnetic (AF) ordering tendency in the staggered periodic Anderson model (PAM) with two alternating inequivalent local moments if their hybridization strengths reside in the Kondo singlet and antiferromagnetic insulator regime separately of the phase diagram of homogeneous PAM. Our results uncover the rich physics induced by the interplay of multiple energy scales in the staggered PAM and furthermore implies the ubiquitous existence of the enhancement of physical quantities in general inhomogeneous systems.
\end{abstract}

\maketitle

\section{Introduction}
Heavy fermion compounds containing one rare-earth ion per unit cell, namely a single unpaired f electron or hole, have been extensively investigated via the theoretical description in terms of Kondo/Anderson lattice models with one spin-1/2 local moment per unit cell~\cite{pam1,pam2,pam3,pam4,pam5,PAMreference}.
The past decade has witnessed the wide demonstration of intriguing phenomena in several families of heavy fermion ternary compounds consisting of multiple inequivalent Kondo sites per unit cell, for instance, two successive magnetic phase transitions in (RE)$_3$Pd$_{20}$X$_6$ (RE = rare earth, X = Si, Ge)~\cite{ternary1,ternary2,ternary3,ternary4,ternary4a}, intermediate valence in Ce-Ru-X (X = In, Al, Sn)~\cite{ternary5,ternary5a}, coexistence of antiferromagnetism and superconductivity in Ce$_3$(Pt/Pd)In$_{11}$~\cite{ternary6,ternary7,ternary8} etc.

These uncovered diverse phenomena call for theoretical investigation of Kondo/Anderson lattices with multiple Kondo ions. Surprisingly, there have been few theoretical studies of this intriguing system of broad interest. 
One important step towards the thorough understanding of the competitive versus cooperative Kondo screening due to multiple local moments was taken in~\cite{Vojta2011}, where a Kondo lattice model (KLM) with two local-moment sublattice coupled to conduction electrons with distinct Kondo couplings is explored in the slave-boson mean-field level. This work has provided a great deal of insights on the Kondo screening and lattice coherence in the heavy Fermi liquid regime of KLM. Another work considered a limiting case, where a sublattice of  f-orbitals in periodic Anderson model (PAM) is depleted, demonstrated some fascinating features associated with a compressible ferrimagnetic ground state~\cite{Costa2018}. However, these previous work either neglected the nonlocal magnetic correlations and only focused on the doped systems~\cite{Vojta2011} or solely took into account the limiting case in the context of understanding the geometry effects of the underlying lattices~\cite{Costa2018}. Even the fundamental questions like the antiferromagnetism has not been fully explored, let alone the exact treatment and exploration of the interplay between multiple energy scales associated with the local moments and Kondo-screened heavy Fermi liquid.

%%%%%%%%%%%%%%%%%%%%%%%%%%%%%%%%%%%%%%%%%%%%%%%%%%%%%%%%%%%%%%%%%%%%%%%%%%
\begin{figure}[b] 
\psfig{figure=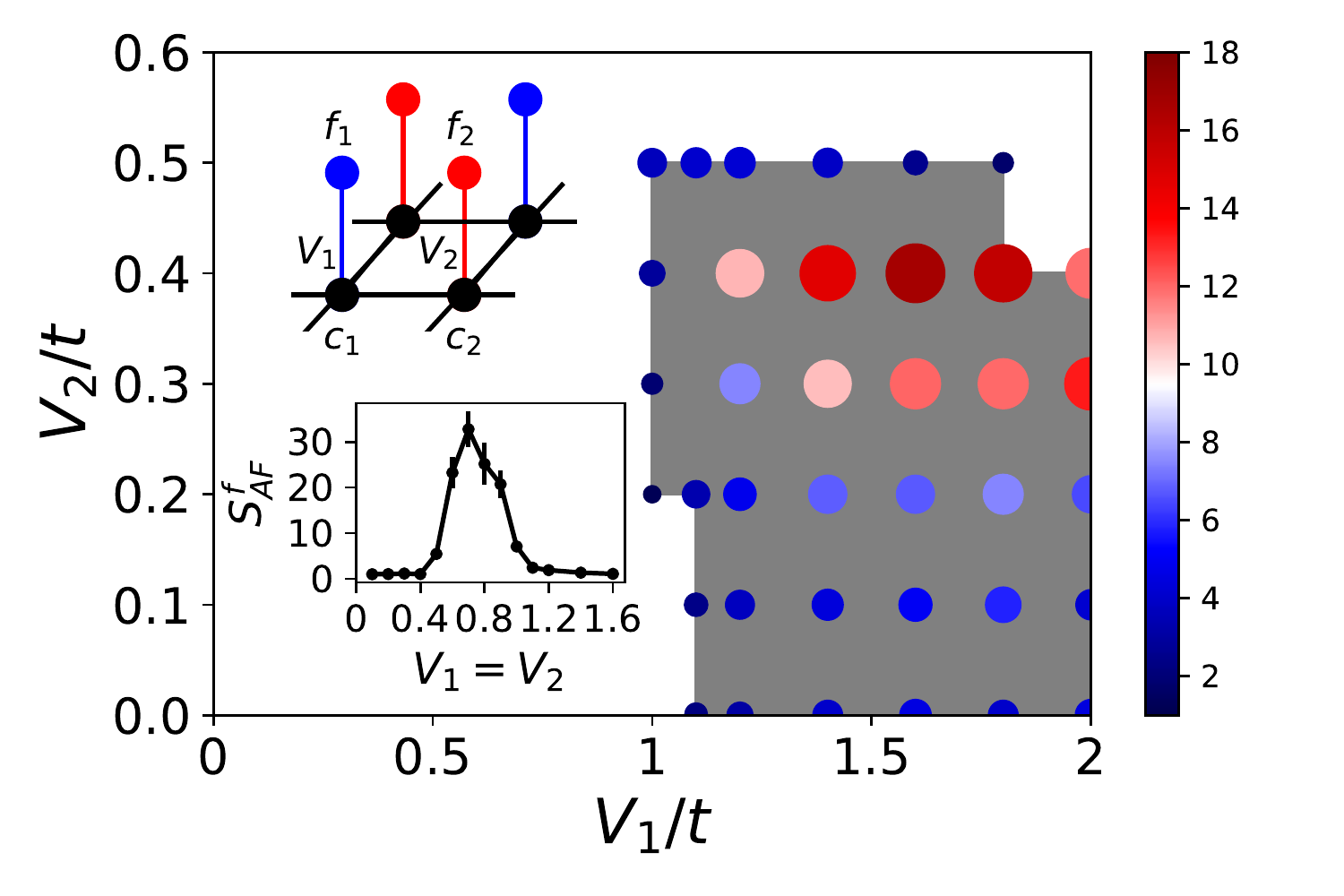}.pdf, height=6.2cm,width=.5\textwidth,angle=0,clip}
\caption{(Color online) Tentative phase diagram of staggered PAM (primitive cell shown as schematic diagram) at lowest simulated temperature $T=0.025t$. The gray regime exhibits enhanced AF ordering tendency compared to the homogeneous counterparts. The enhancement ratio defined in the text is shown via the circle size and colorbar. The white regime denotes the absence of the AF ordering tendency instead of AF order itself. Inset: $S^f_{AF}$ of homogeneous PAM has a peak at $V \sim 0.7t$ facilitates understanding the AF enhancement.}
\label{phase}
\end{figure}

\section{Model and methodology}
To fill this gap, we explored the physics of competing interactions in staggered PAM with two alternating inequivalent local moments on two-dimensional square lattice such that each type of local moment occupies one sublattice (see Figure~\ref{phase}).
Our Hamiltonian in the half-filled form reads:
\begin{eqnarray}
    {\cal H} = &-& t \sum\limits_{\langle ij \rangle \sigma}
(c^{\dagger}_{i\sigma}c_{j\sigma}^{\vphantom{dagger}}
+c^{\dagger}_{j\sigma}c_{i\sigma}^{\vphantom{dagger}}) 
- \sum\limits_{i \sigma}  V_i (c^{\dagger}_{i\sigma}f_{i\sigma}^{\vphantom{dagger}}+ f^{\dagger}_{i\sigma}c_{i\sigma}^{\vphantom{dagger}}) \nonumber \\
    &+& U \sum\limits_{i} (n^{f}_{i\uparrow}-\frac{1}{2}) (n^{f}_{i\downarrow}-\frac{1}{2})
- \mu \sum\limits_{i\sigma} (n^{c}_{i\sigma}+ n^{f}_{i\sigma} )
\label{inPAM}
\end{eqnarray}
where $c^{\dagger}_{i\sigma}(c_{i\sigma}^{\vphantom{dagger}})$
and $f^{\dagger}_{i\sigma}(f_{i\sigma}^{\vphantom{dagger}})$
are creation(destruction) operators for conduction and local electrons on site $i$ with spin $\sigma$.
$n^{c,f}_{i\sigma}$ are the associated number operators.
Here the chemical potential $\mu=0$ assures that both bands are individually half-filled.
$t=1$ is the hopping between conduction electrons on
near neighbor sites $\langle ij \rangle$ and sets the energy scale. $U$ is the local repulsive interaction in the f orbital. The sublattice dependent $V_i=V_1, V_2$ are two distinct hybridizations between conduction electrons and two inequivalent local moments respectively.

It is well known that the essential heavy fermion physics can be captured by the conventional homogeneous PAM at $V_1=V_2$, which exhibits two distinct low temperature phases~\cite{pam1,pam2,pam3,pam4,pam5,PAMreference}. For small $V$, local $f$ moments couple antiferromagnetically via an indirect Ruderman-Kittel-Kasuya-Yosida (RKKY) interaction~\cite{PAMRTS} mediated by the conduction band.  At large $V$, the conduction and local electrons lock into independent singlets so that a paramagnetic spin liquid ground state forms. This reflects a competition between the RKKY and Kondo energy scales, $\sim J^{2}/W$ and $\sim W e^{-W/J}$, respectively, with $J\sim V^{2}/U_f$ and $W$ the bandwidth.
It is natural to ask whether the two distinct Kondo screening energy scales due to inequivalent local moments might render the staggered PAM some new physics, for which some general insightful discussion has been put forward by previous work~\cite{Vojta2011}. 
Here we will focus on the staggered PAM at half-filling and point out that even this limiting case brings about strikingly new phenomenon in terms of a remarkable enhancement of the lattice antiferromagnetism. In the absence of Hubbard interaction, the two-site two-orbital unit cell (see Fig.~\ref{phase}) gives rise to four energy bands $E(\mathbf{k})$ satisfying $E^2(\mathbf{k})=\frac{1}{2} (A \pm \sqrt{A^2-4V^2_1 V^2_2})$ with $A \equiv \epsilon^2_{\mathbf{k}}+V^2_1+V^2_2$ as the generalization of homogeneous PAM, so that the system is readily a band insulator for any finite $V_{1,2}$ at half-filling.

To treat with the Hubbard interaction together with diverse energy scales associated with two inequivalent local moments on the equal footing, we solve Eq.~\ref{inPAM} by means of the finite temperature determinant Quantum Monte Carlo (DQMC)~\cite{blankenbecler81}, where a path integral expression is written for the quantum partition function ${\cal Z}={\rm Tr\,exp}\,(\,-\beta {\cal H} \,)$. The Hubbard interaction term for local moments is mapped onto a coupling of the $f$ electron spin with a space and imaginary-time dependent auxiliary (``Hubbard-Stratonovich'') field. In this manner, the interaction energy is treated without approximation \cite{trotter} and the fermionic degrees of freedom can be integrated out analytically, which results in an exact expression for ${\cal Z}$ and operator expectation values for various physical correlation functions in terms of integrals over the field configurations $\{ S_{i\tau} \}$, whose sampling is realized in a Monte Carlo algorithm. The half-filled case of Eq.~\ref{inPAM} ensures the absence of infamous Fermionic sign problem so that the physical quantities can be evaluated at low enough temperatures~\cite{loh90}.

Without loss of generality, we assume that $V_1 \geq V_2$ and concentrate on the characteristic intermediate coupling strength $U=4.0t$, which is commonly used in the literature to explore the essential physics within PAM. Another reason lies that it has been well established that the critical $c-f$ hybridization strength separating the Kondo singlet and antiferromagnetic insulating ground states is $V_c \sim 1.0t$~\cite{wenjian,note}. 
Besides, to study a large enough lattice at low enough temperature with manageable computational cost, most results presented are for imaginary-time
interval $d\tau=0.125$\cite{trotter} and $12 \times 12$ lattices with periodic boundary~\cite{notefs}. 
The physical quantity we focus on is the antiferromagnetic (AF) structure factor for f-orbital local moments $S^f_{AF}(V_1,V_2)=\frac{1}{N} \sum_{ij} e^{-i \mathbf{q} \cdot (\mathbf{R_i}-\mathbf{R_j})} \langle (n^f_{i\uparrow}-n^f_{i\downarrow}) (n^f_{j\uparrow}-n^f_{j\downarrow}) \rangle$ at $\mathbf{q}=(\pi,\pi)$, where $\mathbf{R_i}$ denotes the coordinates of site $i$ and $N$ is the lattice size. Note that the perfect AF order corresponds to $S^f_{AF}=N$ so that it diverges below the Neel temperature in the thermodynamic limit.

%%%%%%%%%%%%%%%%%%%%%%%%%%%%%%%%%%%%%%%%%%%%%%%%%%%%%%%%%%%%%%%%%%%%%%%%%%%
\section{Results}\label{results}
\begin{figure*}[t] 
\psfig{figure=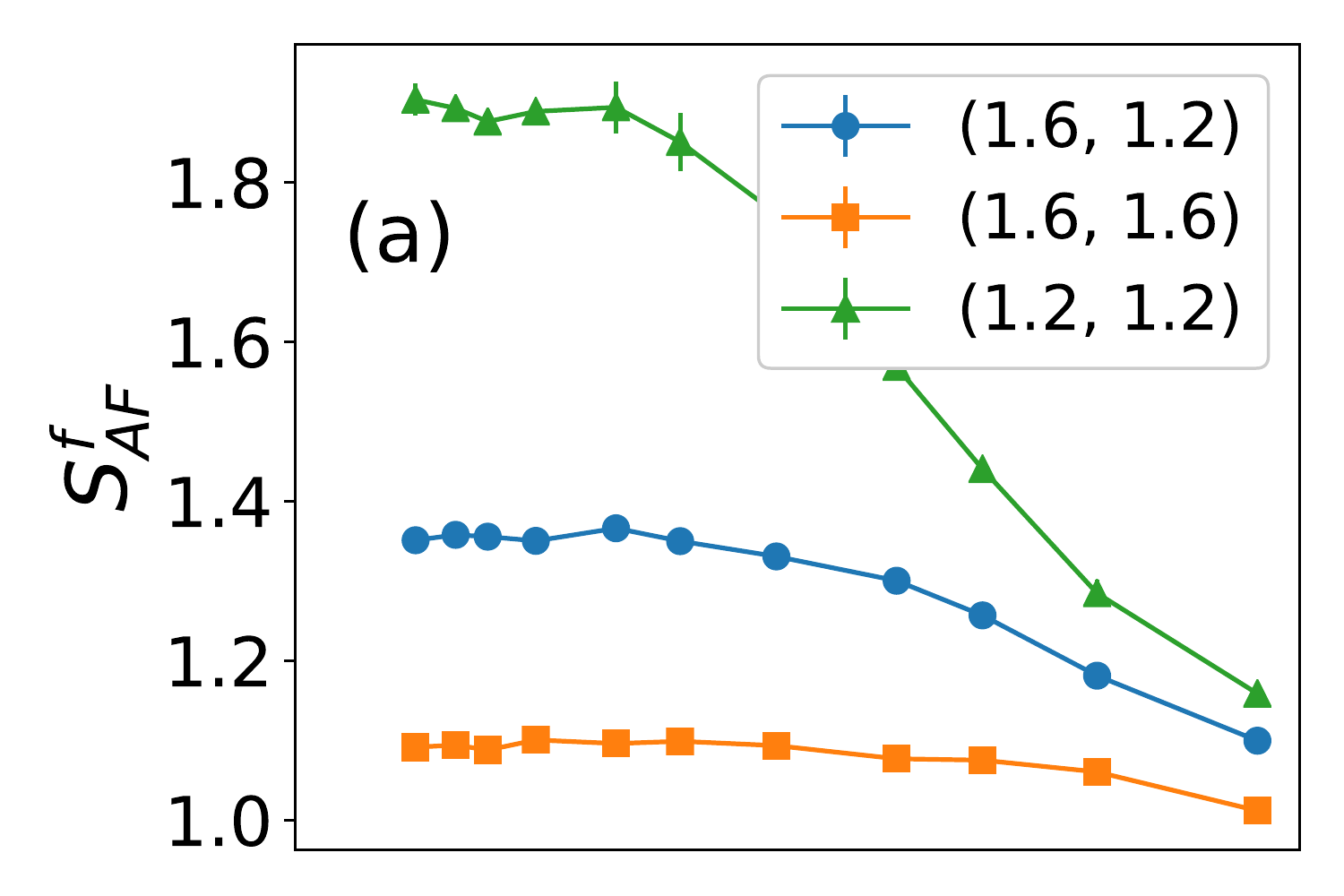}.pdf,
width=.32\textwidth,angle=0,clip}
\psfig{figure=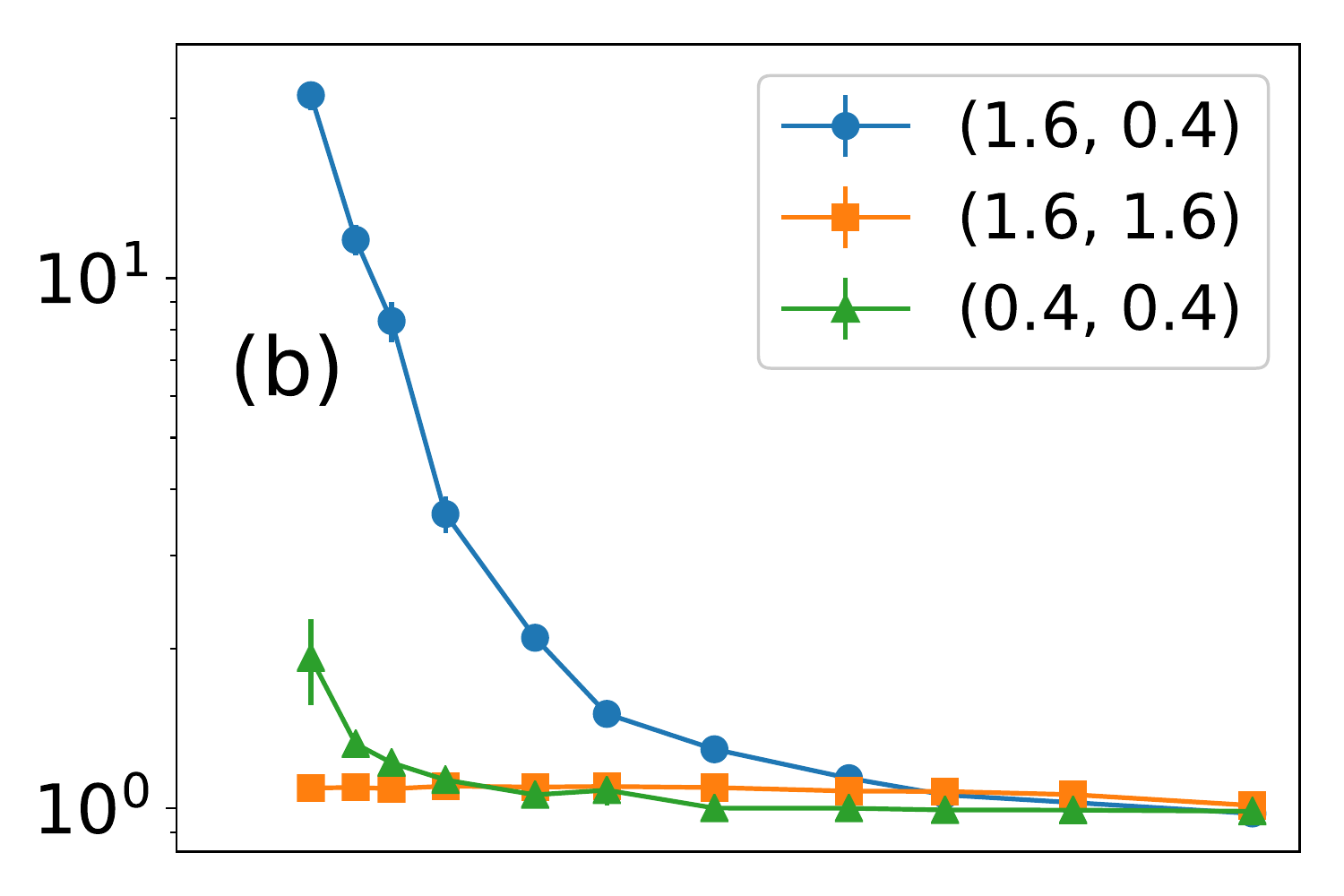}.pdf,
width=.32\textwidth,angle=0,clip} 
\psfig{figure=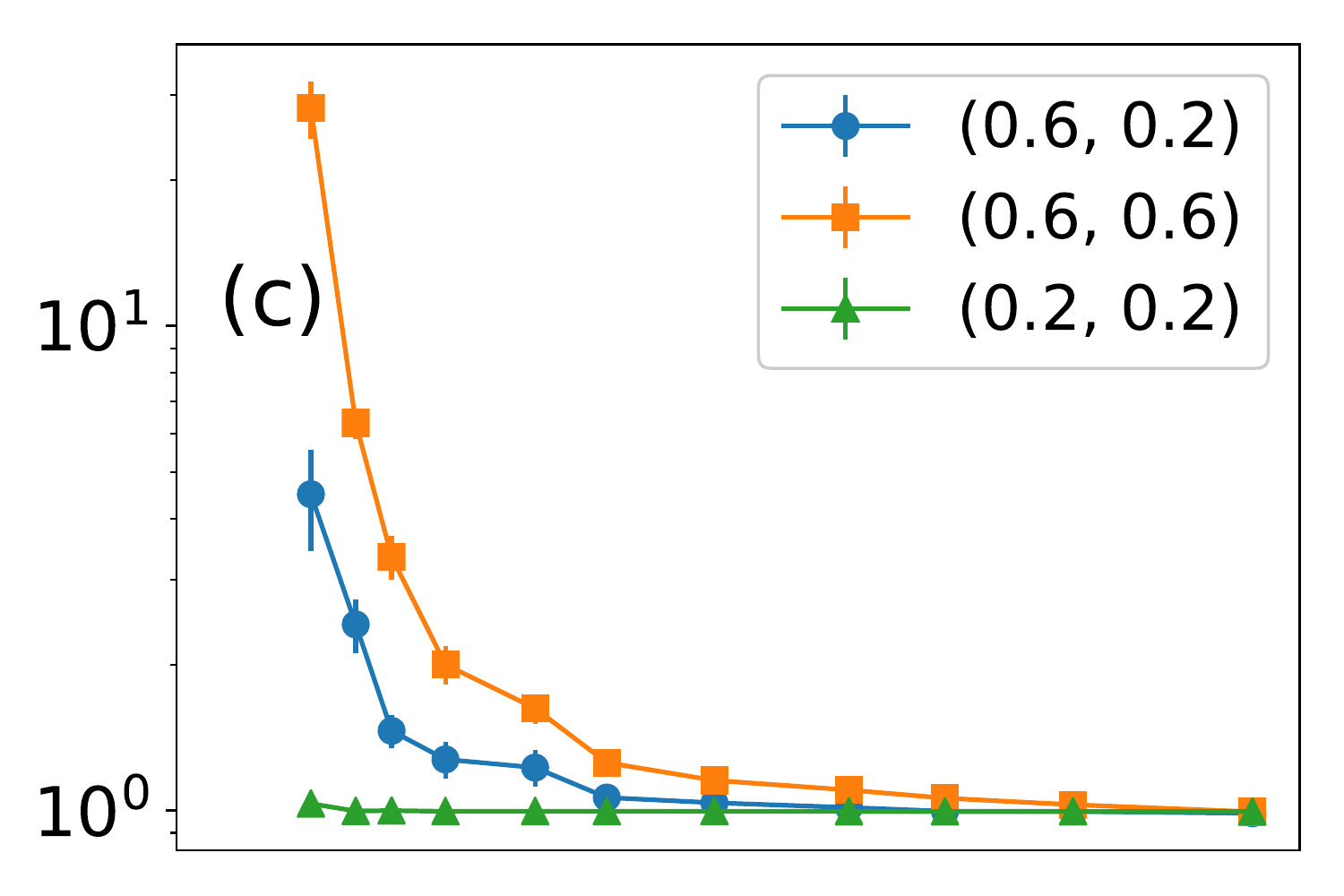}.pdf,
width=.32\textwidth,angle=0,clip} 
\psfig{figure=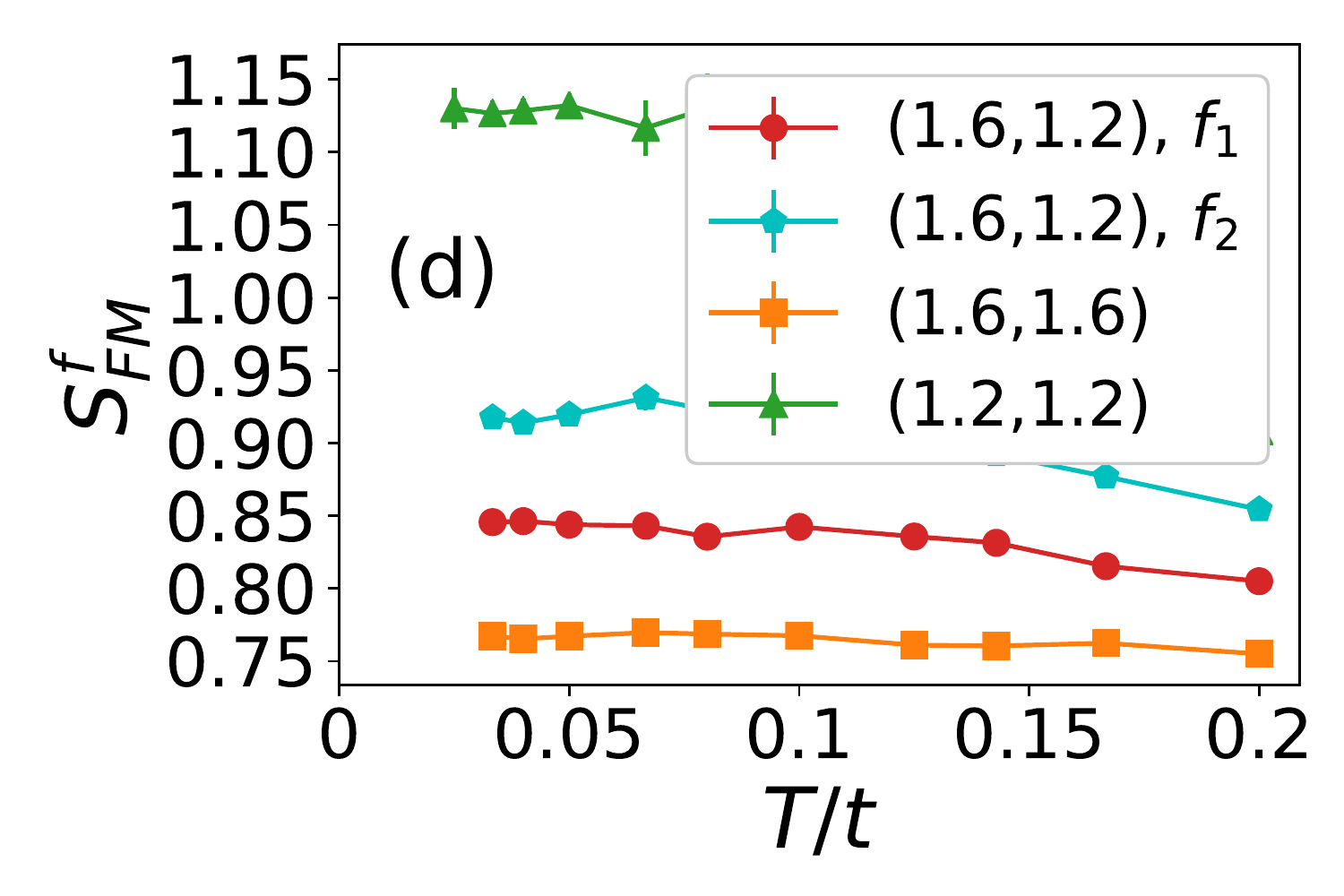}.pdf,
width=.32\textwidth,angle=0,clip} 
\psfig{figure=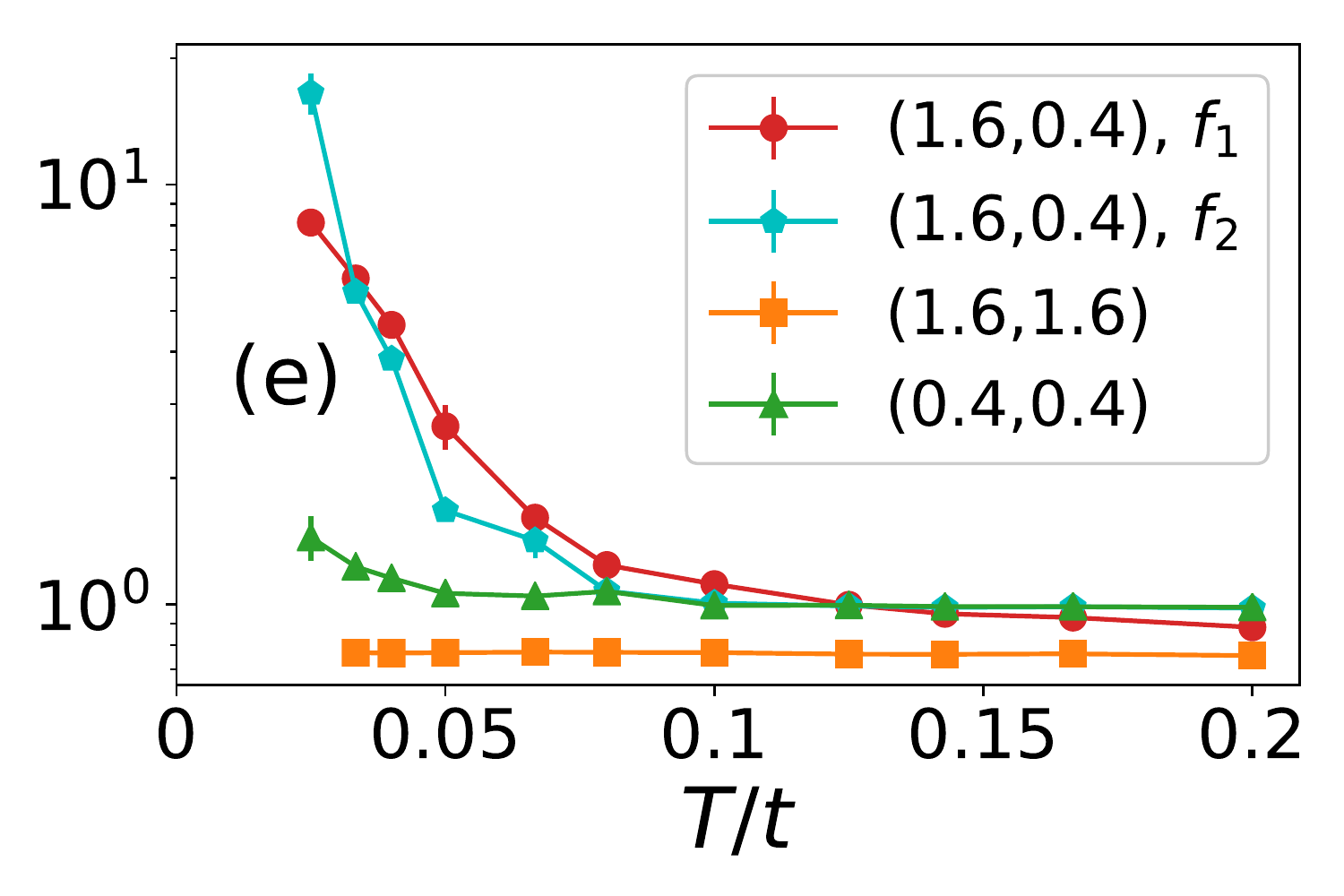}.pdf,
width=.32\textwidth,angle=0,clip} 
\psfig{figure=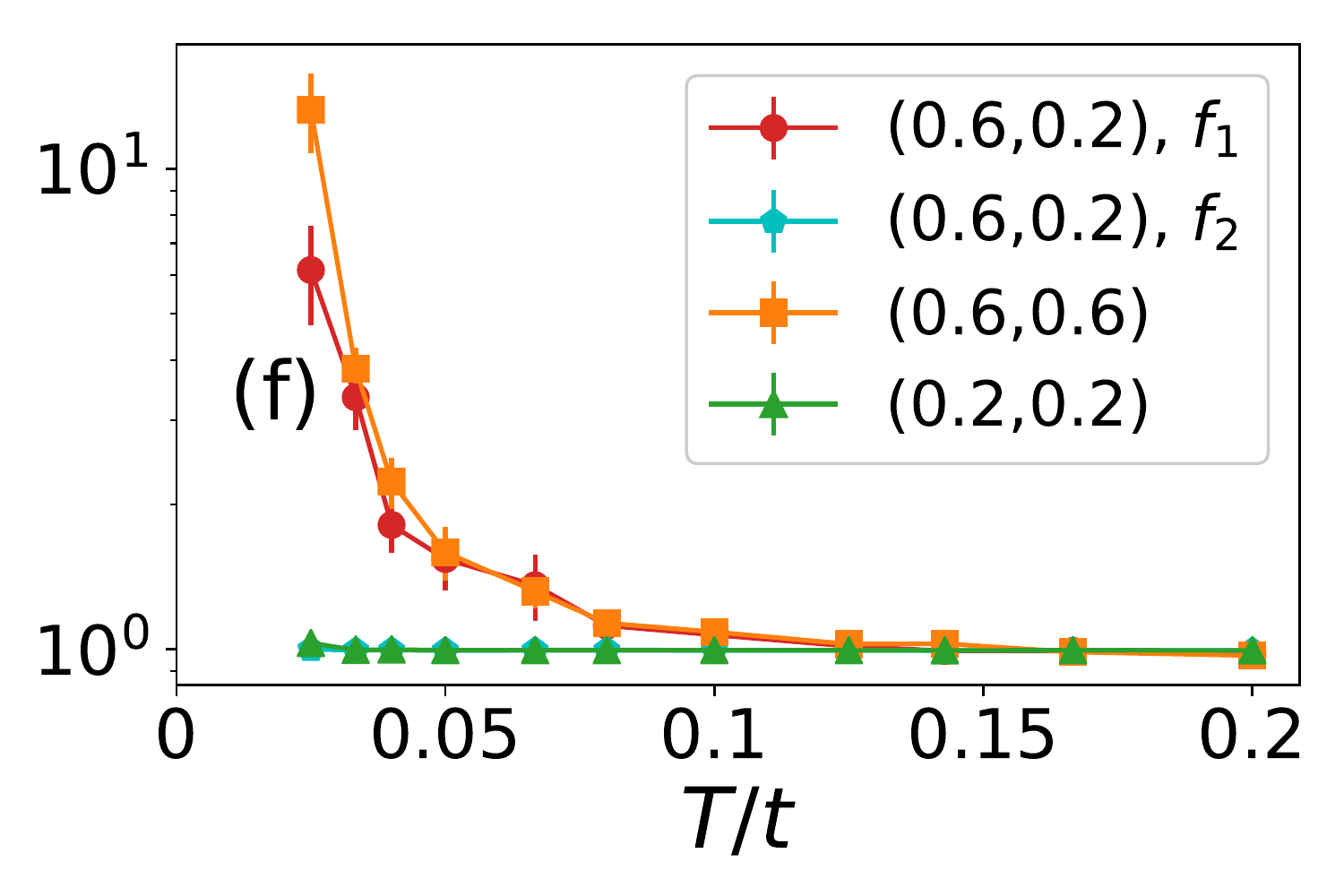}.pdf,
width=.32\textwidth,angle=0,clip}
\caption{(Color online) Temperature evolution of (a-c) $S^f_{AF}$ and (d-f) $S^f_{FM}$ of two sublattice corresponding to $f_{1,2}$ separately in three characteristic systems with distinct behavior of $S^f_{AF}$ compared with their counterparts in homogeneous PAM.}
\label{SafT}
\end{figure*}

Our major finding is illustrated as the tentative phase diagram at our lowest simulated temperature $T=0.025t$ displayed in Fig.~\ref{phase}. The gray regime highlighted exhibits remarkable enhancement of $S^f_{AF}(V_1,V_2)$ in the staggered PAM compared to its counterparts $S^f_{AF}(V_1)$ and $S^f_{AF}(V_2)$ in the homogeneous systems. 
Clearly, the AF enhancement requires that $V_1$ resides in the Kondo singlet regime of homogeneous PAM. Meanwhile, $V_2$ is restricted up to $\sim 0.5t$ in the AF insulating regime of standard PAM.
Moreover, the enhancement ratio $r$, defined as $S^f_{AF}(V_1,V_2)/\max\{S^f_{AF}(V_1),S^f_{AF}(V_2)\}$ and shown via the circle size and colorbar in Fig.~\ref{phase}, exhibits nontrivial dependence on $(V_1,V_2)$. In particular, $r$ can be as large as one order of magnitude within $V_2=0.3-0.4t$ peaked at $V_1=1.6t, V_2=0.4t$ and decreases when $V_1$ and/or $V_2$ leave away from this parameter regime. Note that $V_2=0$ axis reduces to the depleted PAM~\cite{Costa2018}, where our antiferromagnetism recovers the previously observed ferrimagnetism due to the absence of the coupling between two distinct local moments.

One important issue we remark here is that the gray regime in Fig.~\ref{phase} exhibiting the AF enhancement precisely implies the enhanced AF ordering tendency but does not necessarily mean that this regime can or the unlabeled white regime cannot definitely host the AF order, whose existence has to be decisively determined by more thorough investigation at zero temperature in the thermodynamic limit that is out of scope in this work.

\begin{figure*}[t] 
\psfig{figure=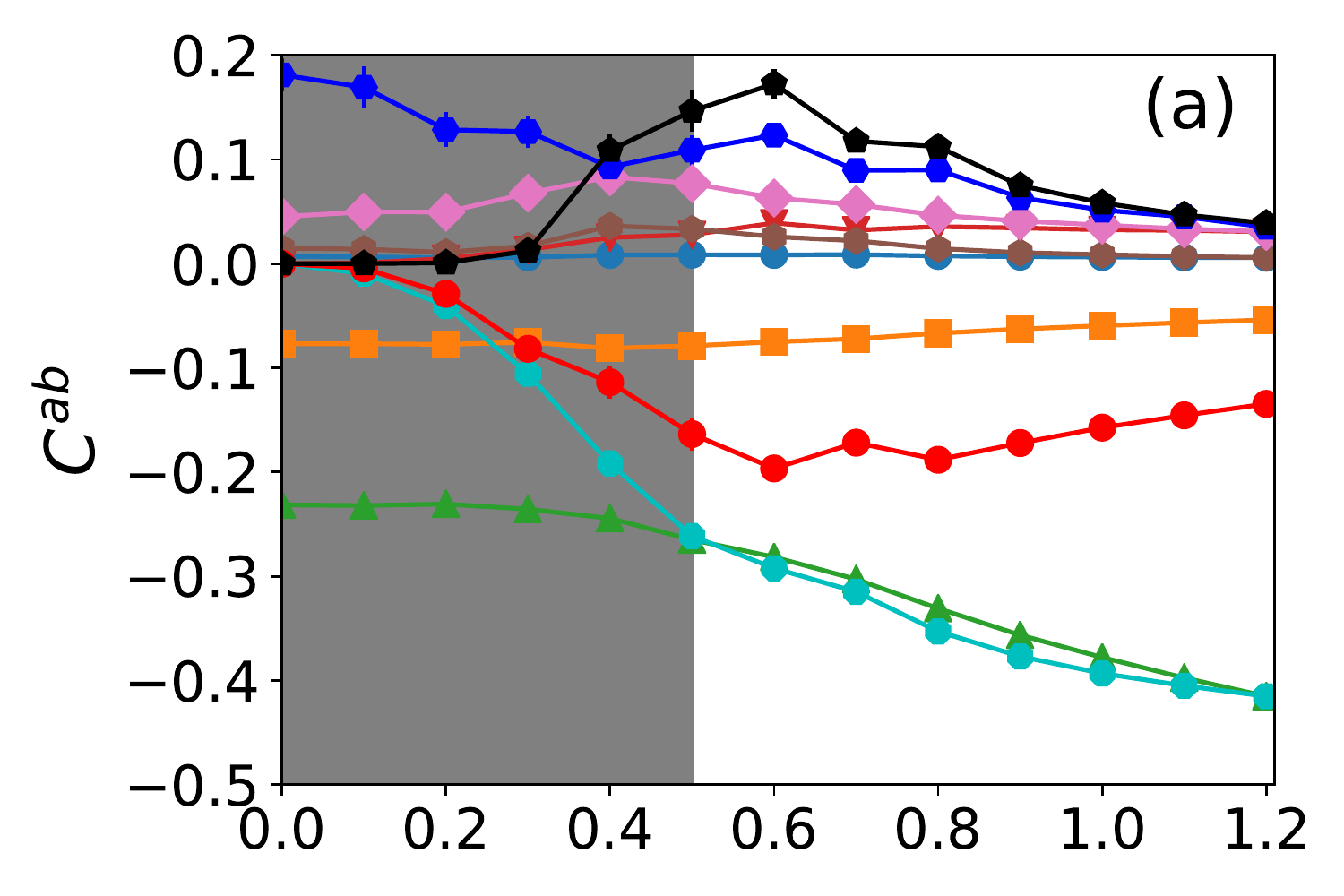}.pdf,width=.45\textwidth,angle=0,clip}%=true, trim = 0.1cm 0.5cm 0.4cm 0.0cm}
\psfig{figure=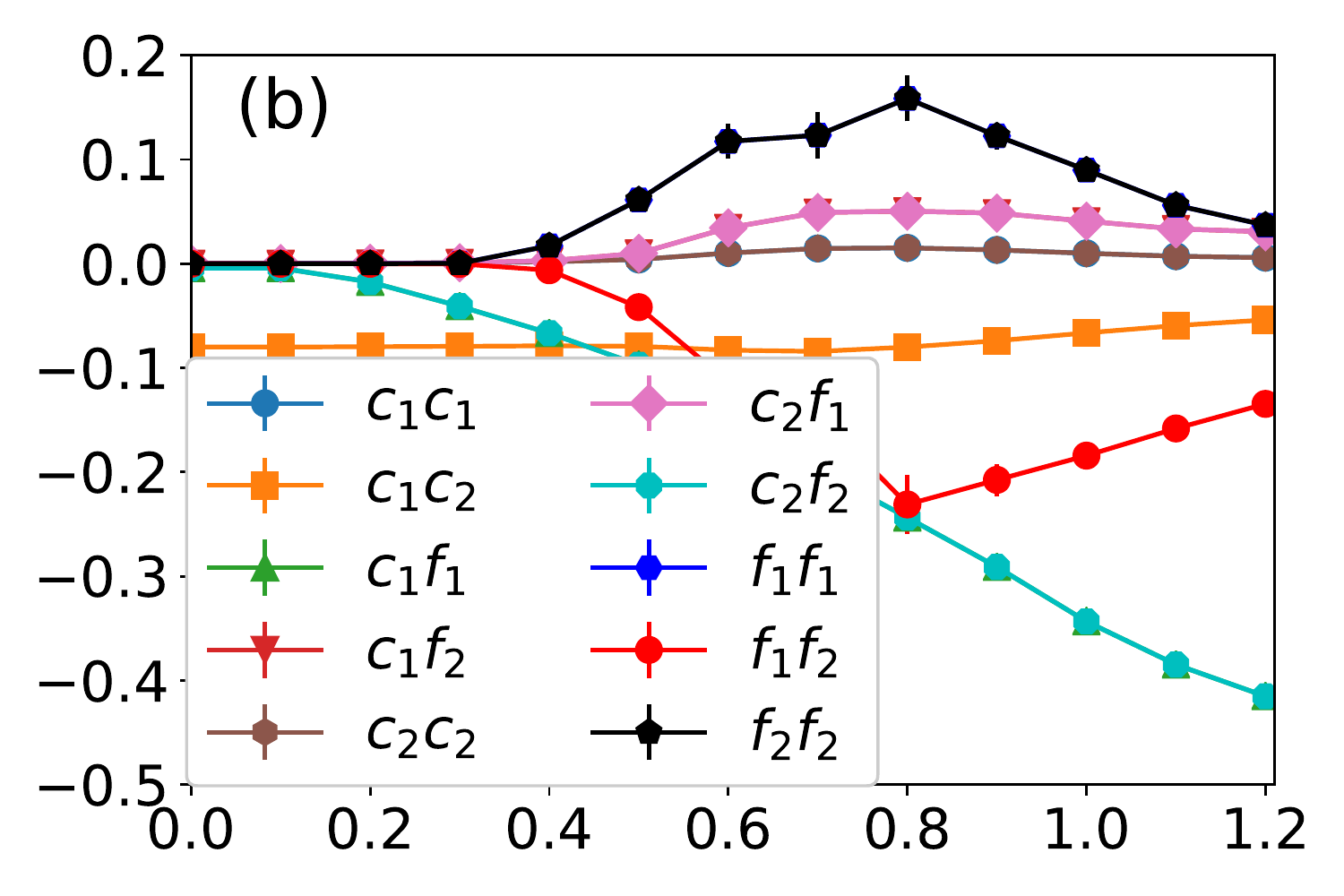}.pdf,width=.45\textwidth,angle=0,clip} \\%=true, trim = 0.1cm 0.5cm 0.4cm 0.0cm}
\psfig{figure=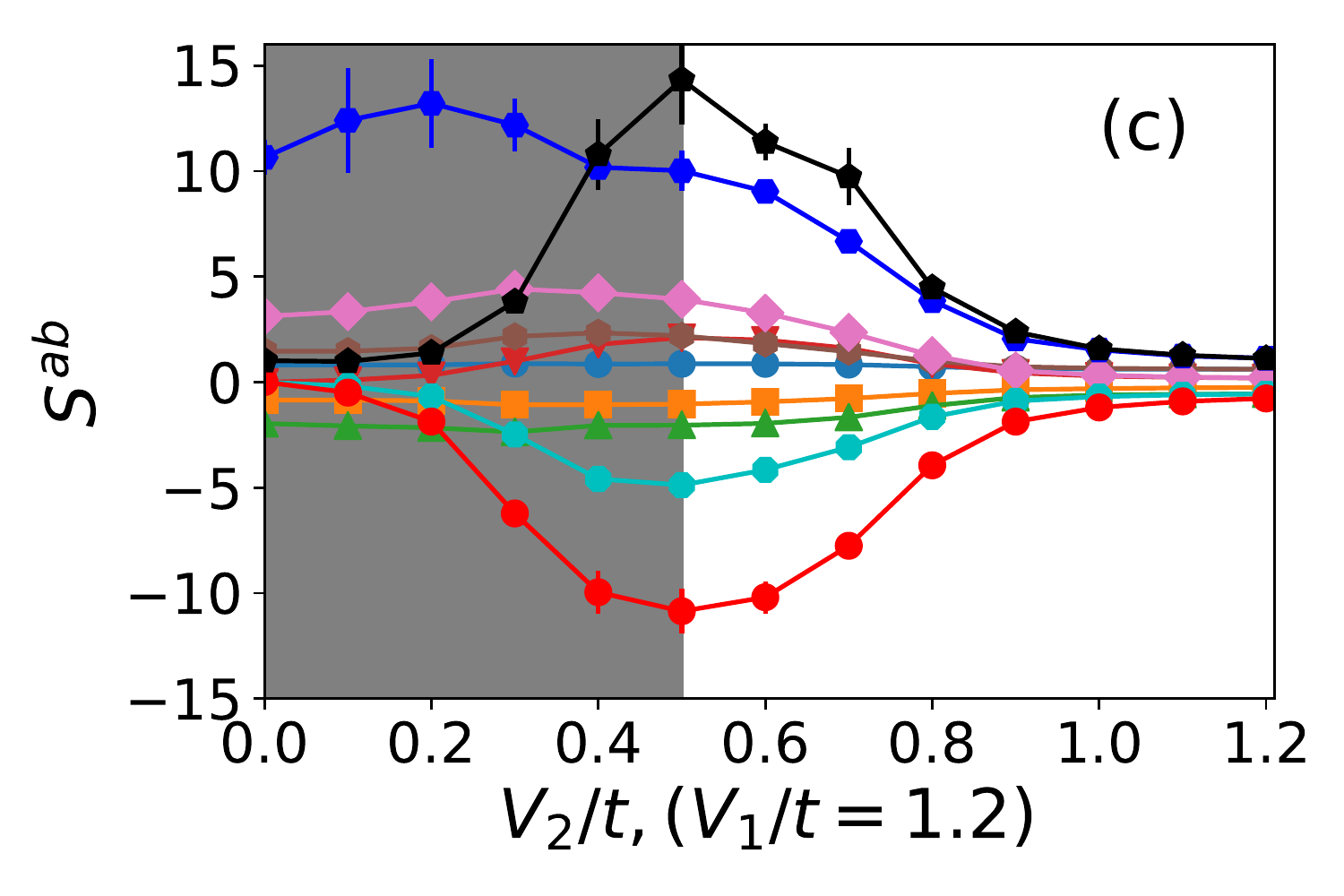}.pdf, width=.45\textwidth,angle=0,clip}
\psfig{figure=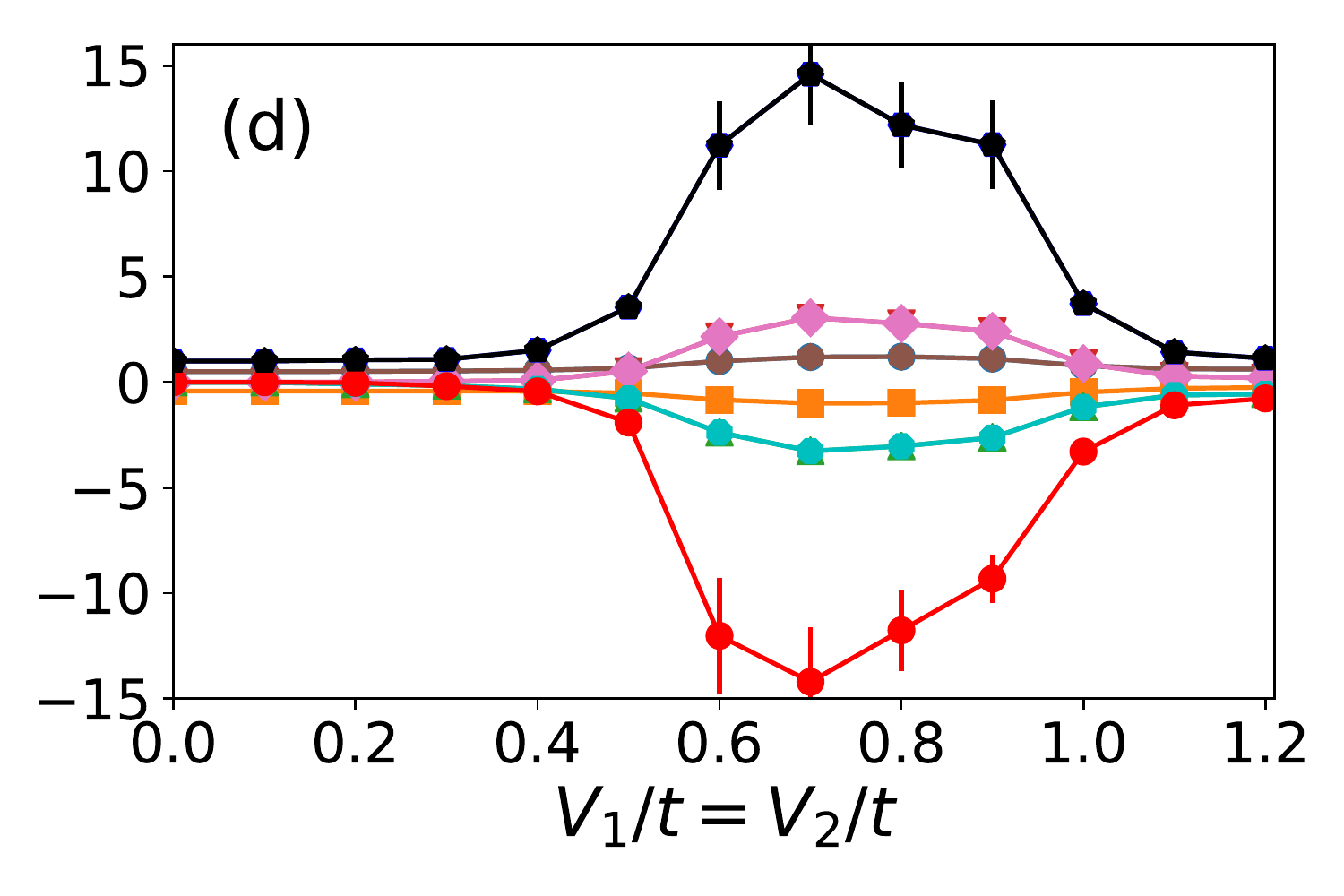}.pdf, width=.45\textwidth,angle=0,clip}
\caption{(Color online) Individual (a-b) nearest-neighbor spin correlations $C^{ab}$ and (c-d) structure factors $S^{ab}$ between $c_1,c_2,f_1,f_2$ orbitals of systems with fixed $V_1=1.2t$ and varied $V_2$ compared with their counterparts in homogeneous systems (b, d) at lowest accessed $T=0.025t$. The gray color highlights the AF enhancement regime.}
\label{vsV2}
\end{figure*}

These features can be understood qualitatively by $S^f_{AF}(V)$ in the homogeneous PAM as shown in the inset, which indicates that the maximum $S^f_{AF}(V)$ occurs at $V \sim 0.7t$ due to the competition between RKKY interaction mediated by $c-f$ hybridization $V$ and Kondo hybridization~\cite{AFnote}. 
Generally, if two sublattice with distinct local moments are nested with each other to form a staggered PAM, there is a ``self-averaging'' effect resulting the reduced (enhanced) effective $V_1(V_2)$. In fact, our numerical evidence of the two local moments (not shown) and $c-f$ spin correlations (Fig.~\ref{vsV2}(a-b)) support this scenario. Based on this common effect, if $S^f_{AF}(V)$ varies monotonically between $V=V_{1,2}$, as exemplified in Fig.~\ref{SafT}(a,c), $S^f_{AF}(V_1,V_2)$ lies between $S^f_{AF}(V_1)$ and $S^f_{AF}(V_2)$ as expected. Nonetheless, a nontrivial dependence of $S^f_{AF}(V)$ peaked at $V=0.7t$ results in the possible AF enhancement if $V_{1,2} \rightarrow 0.7t$ from two sides. Besides, it also implies the difficulty, if not impossible due to our finite temperature simulations, of observing the AF enhancement if $V_{1,2}$ are already near $V=0.7t$, e.g. $(V_1,V_2)=(0.9,0.6)$ with significantly large $S^f_{AF}(V)$ at $V=0.6-0.9t$ of homogeneous PAM.

More insight on the AF enhancement can be gathered by temperature effects. To this aim, Figure~\ref{SafT} (a-c) illustrates the temperature evolution of $S^f_{AF}$ in three characteristic systems with distinct behavior of $S^f_{AF}(V_1,V_2)$. Obviously, if $V_{1,2}$ both exceed $V_c=1.0t$ or not of homogeneous PAM, namely both in the Kondo singlet or AF insulator regime, $S^f_{AF}(V_1,T)<S^f_{AF}(V_1,V_2,T)<S^f_{AF}(V_2,T)$ in the intermediate temperature regime shown here by the ``self-averaging'' effect discussed previously. In contrary, the distinct behavior at $(V_1,V_2)=(1.6,0.4)$ (Fig.~\ref{SafT} (b)) shows a striking enhancement in a wide temperature regime and the enhancement ratio is as large as one order of magnitude. A closer examination indicates that the enhancement starts from a characteristic high temperature scale. Furthermore, we remark that even though there is no enhancement in some cases e.g. at $V_1=1.2t, V_2=0.6t$ at lowest temperature $T=0.025t$, a considerable enhancement can still be observable in an intermediate temperature regime until a lower temperature scale at which the enhancement disappears. 

Until now we have only concentrated on the global lattice quantity $S^f_{AF}$. It is natural to ask which components mostly contribute to the AF enhancement. To this aim, we define the individual structure factor between different orbitals $S^{ab}=\frac{1}{N} \sum_{ij} e^{-i \mathbf{q} \cdot (\mathbf{R_i}-\mathbf{R_j})} \langle (n^a_{i\uparrow}-n^a_{i\downarrow}) (n^b_{j\uparrow}-n^b_{j\downarrow}) \rangle$ at $\mathbf{q}=(\pi,\pi)$ with $a,b=c_1,c_2,f_1,f_2$ shown in Fig.~\ref{phase}. 
Note that we will reverse the sign of $S^{ab}$ if it is negative to be consistent with the sign of the local spin correlations defined later, which is equivalent to define $S^{ab}$ at $\mathbf{q}=0$.
In this way $S^f_{AF}=S^{f_1f_1}+S^{f_2f_2}-2S^{f_1f_2}$ and $S^{f_{i}}_{FM}=S^{f_if_i}, i=1,2$. 
Fig.~\ref{SafT} (d-f) illustrates the temperature evolution of the ferromagnetic structure factor $S^f_{FM}$ for two sublattice $f_{1,2}$ respectively compared with their counterparts in homogeneous PAM with $V_{1,2}$. The ``self-averaging'' induced approaching of two sublattice $S^f_{FM}$ is clearly visible in all cases despite that the presence of AF enhancement (b) naturally coincides with the FM enhancement of two sublattice (e). In addition, in the case of $V_1=1.6t,V_2=0.4t$, $S^f_{FM}$ of $f_{1,2}$ sublattice have the same scale without dominance of any one, although the $f_2$ sublattice with smaller hybridization $V_2$ exceeds that of $f_1$ sublattice as expected from the perspective of homogeneous PAM.  

To further reveal the distinct low temperature properties of the staggered PAM and the relative importance of $f_1$ or $f_2$ sublattice, we examine the individual contributions of all four orbitals of systems with fixed $V_1=1.2t$ and varied $V_2$ at $T=0.025t$ shown in Figure~\ref{vsV2}(a,c) compared with their counterparts in homogeneous systems shown in Figure~\ref{vsV2}(b,d). The gray color highlights the AF enhancement regime. As shown in Fig.~\ref{vsV2}(a-b), we first focus on the locally nearest-neighbor spin correlations $C^{ab}=\langle (n^a_{\uparrow}-n^a_{\downarrow}) (n^b_{\uparrow}-n^b_{\downarrow}) \rangle$ corresponding to $S^{ab}$ defined earlier, which is consistent with Shen's theorem~\cite{shunqing} that $C^{ab}$ is positive (negative) for site/orbital on the same (different) sublattice of bipartite lattices.

As shown in Fig.~\ref{vsV2}(a-b), the significant difference between staggered and homogeneous PAM is manifested via the much stronger $f_1f_1$ (blue), $c_1f_1$ (green), $f_1f_2$ (red) spin correlations~\cite{notecf} in staggered PAM at $V_2<0.5t$, which strongly indicates the vital role of $f_1$ local moments with larger hybridization $V_1$, despite of the rapid growth of $f_2f_2,c_2f_2$ components starting from $V_2=0.3t$. Simultaneously, due to the coupling between various components, $c_2f_1$ has minor increase accordingly. Note that the bifurcation between $c_1f_1$ and $c_2f_2$ components at the crossover point $V_2=0.5t$ is consistent with Fig~\ref{phase}. Therefore, we remark that our AF enhancement mechanism is distinctly different from the ferrimagnetism induced by noninteracting orbitals, where the $c_2c_2$ correlation should be enhanced~\cite{Costa2018}.

The analysis of the short range spin correlations $C^{ab}$ between different orbitals provided some insight into the origin of the AF enhancement. More evidence on the decisive role of $f_1$ local moments can be provided via the individual structure factor $S^{ab}$. Fig.~\ref{vsV2}(c-d) demonstrates all individual contributions that are essentially similar to the behavior of $C^{ab}$ in the regime of $V_2<0.5t$ hosting the AF enhancement. Apparently, the inclusion of longer range spin correlations induces the diminish of $S^{c_if_i},i=1,2$ at sufficiently large $V_2$. Most importantly, the dominance of $S^{f_1f_1}$ over $S^{f_2f_2}$ at weak $V_2<0.4t$ is gradually replaced by the rapid increase of $S^{f_2f_2}$ with the same scale as $S^{f_1f_1}$ starting from $V_2=0.3t$, which results in the large AF enhancement, although the enhancement ratio decreases (see Fig.~\ref{phase}) due to the simultaneous increase of $S^f_{AF}$ in the corresponding homogeneous systems.

Finally, one may notice that the staggered pattern of two inequivalent local moments is exactly commensurate with the antiferromagnetic wave vector $(\pi,\pi)$ where the $S_{AF}$ has considerable enhancement. We point out that this is essential to realize the AF enhancement. In fact, we confirm that a similar enhancement in $(\pi,0)$ wave vector does not occur if two inequivalent local moments have a stripe-like pattern with $(\pi,0)$ modulation.

\section{Summary and outlook}
To summarize, we performed a determinant QMC calculation of the staggered PAM with two alternating inequivalent local moments to demonstrate that the lattice antiferromagnetic ordering tendency can be strikingly enhanced if the hybridization strengths of two local moments with the conduction electrons reside in the Kondo singlet and antiferromagnetism regime separately of the phase diagram of the homogeneous PAM. In other words, the antiferromagnetic correlation can be largely enhanced via replacing a sublattice of local moments by singlets with much stronger $c-f$ hybridization. 
We emphasized that this AF enhancement precisely implies the enhanced AF ordering tendency but does not necessarily mean that the corresponding parameter regime in our phase diagram Fig.~\ref{phase} can deterministically host the AF order, whose existence has to be decisively determined by more thorough investigation at zero temperature in the thermodynamic limit that is out of scope in this work.
This counterintuitively new route of strengthening the AF order can be interpreted as the consequence of the ``self-averaging'' effect between two distinct local moments together with the non-monotonic dependence of $S^f_{AF}$ in homogeneous PAM.

The work presented here indicates that the interplay of multiple energy scales corresponding to two inequivalent Kondo screenings and antiferromagnetism in a single system like staggered PAM deserves deeper understanding through more extensive investigation. 
In addition, it is requisite to further explore the connection of our findings to the realistic heavy fermion materials e.g. Ce$_3$(Pt/Pd)In$_{11}$~\cite{ternary6,ternary7,ternary8} that has layered structure where two inequivalent Ce ions locate on different layers, which might challenge our model of two ions intertwined in a single layer.
Moreover, the ubiquitous existence of the two essential ingredients, namely the ``self-averaging'' effects and non-monotonic dependence of physical quantites in homogeneous system, suggests the probable generality of the enhancement of particular physical properties in other systems. For example, provided that there has been increasing evidence supporting an underlying connection between the Hubbard model and PAM~\cite{Held2000,yifeng}, we conjecture about a similar AF enhancement in the conventional Hubbard model with staggered interaction strengthes.

Furthermore, similar to the doped Hubbard model as the standard framework of cuprate physics, the doping effect of the staggered PAM as a relevant fundamental model of the family of heavy fermion compounds~\cite{ternary6,ternary7,ternary8} and accordingly the competition and/or coexistence between the enhanced AF and superconductivity~\cite{WeiWu} would be highly interesting. The investigation of these intriguing questions are in progress.

\begin{acknowledgments}
 {\em Acknowledgements:} We thank Richard Scalettar, Mona Berciu, Yi-feng Yang, and Natanael Costa for useful discussions. This work was funded by the Stewart Blusson Quantum Matter Institute at University of British Columbia, and by the Natural Sciences and Engineering Research Council of Canada.
\end{acknowledgments}

%%%%%%%%%%%%%%%%%%%%%%%%%%%%%%%%%%%%%%%%%%%%%%%%%%%%%%%%%%%%%%%%%%%%%%%%%%%


\begin{thebibliography}{27}
\bibitem{trotter}
The Trotter error associated with the discretization of inverse
temperature is typically smaller than statistical
errors from the Monte Carlo sampling, and
can be eliminated by extrapolation.
\bibitem{note}
Note that Ref.~\cite{pam5} recently reported that the quantum critical hybridization separating zero temperature antiferromagnetic insulator and a Kondo insulator of the two-dimensional PAM lies at $V_c/t=0.91$.
\bibitem{notefs}
We have numerically checked the robustness of AF enhancement in varied lattice sizes.
\bibitem{AFnote}
Our $S^f_{AF}$ at the reported quantum critical hybridization $V_c=1.0t$~\cite{wenjian} or $V=0.91t$~\cite{pam5} is not zero due to the nature of our finite temperature simulations on finite lattices.
\bibitem{notecf}
Note that in the homogeneous PAM shown in Fig.~\ref{vsV2}(b-d), $C^{c_1f_1}$ (green) is the same as $C^{c_2f_2}$ (cyan). Similarly, $S^{c_1f_1}=S^{c_2f_2}$.
\end{thebibliography}
\end{document}